\documentclass[12pt]{article}
\usepackage{epsf} 
\usepackage{amsmath}
\usepackage{amssymb}
\usepackage{epsfig}
\usepackage{latexsym}
\usepackage{amsfonts}
\usepackage{graphicx}%
\usepackage{varioref}
\usepackage{ifthen}
\begin{document}
\parindent 0mm 
\setlength{\parskip}{\baselineskip} 
\thispagestyle{empty}
\pagenumbering{arabic} 
\setcounter{page}{0}
\mbox{ }
\rightline{UCT-TP-270/07, MZ-TH/07-18}
%\newline
%\rightline{MZ-TH/07-18}
\newline
\rightline{December 2007}
\newline
\rightline{revised April 2008}
\newline
%\vspace{0.1cm}
\begin{center}
{\large {\bf 	Strange quark   mass from Finite Energy QCD sum rules to five loops}}
{\LARGE \footnote{{\LARGE {\footnotesize Supported in part by  NRF (South Africa) and DFG (Germany).}}}}
\end{center}
%\vspace{.05cm}
\begin{center}
{\bf Cesareo A. Dominguez}$^{(a)}$, {\bf Nasrallah F. Nasrallah}$^{(b)}$, {\bf Raoul R\"{o}ntsch}$^{(a)}$, {\bf Karl Schilcher}$^{(c)}$
\end{center}
\begin{center}
$^{(a)}$Centre for Theoretical Physics and Astrophysics\\[0pt]University of
Cape Town, Rondebosch 7700, South Africa\\
$^{(b)}$ Faculty of Science, Lebanese University, Tripoli, Lebanon\\
$^{(c)}$ Institut f\"{u}r Physik, Johannes Gutenberg-Universit\"{a}t\\
Staudingerweg 7, D-55099 Mainz, Germany
\end{center}
%\vspace{0.2cm}
\begin{center}
\textbf{Abstract}
\end{center}
\noindent
The strange quark mass is determined from a new QCD Finite Energy Sum Rule (FESR) optimized to reduce considerably the systematic uncertainties arising from the  hadronic resonance sector. As a result, the main uncertainty in this determination is due to the value of $\Lambda_{QCD}$. The correlator of axial-vector divergences is used in perturbative QCD to five-loop order, including quark and gluon condensate contributions, in the framework of both Fixed Order (FOPT), and Contour Improved Perturbation Theory (CIPT). The latter exhibits very good convergence, leading to a remarkably stable result in the very wide range $s_0 = 1.0 - 4.0 \;\mbox{GeV}^2$, where $s_0$ is the radius of the integration contour in the complex energy (squared) plane. The value of the strange quark mass in this framework at a scale of 2 GeV is $m_s(2\; \mbox{GeV}) = 95 \pm 5\,(111 \pm 6) \;\mbox{MeV}$ for $\Lambda_{QCD} = 420 \,(330) \; \mbox{MeV}$, respectively.\\


\newpage
\bigskip
\noindent
\section{Introduction}
\noindent
The strange quark mass is a very important QCD parameter measuring the strength of chiral $SU(3) \otimes SU(3)$, and flavour $SU(3)$ symmetry breaking. It also has a strong impact on a variety of QCD applications in weak hadronic physics. For this reason, many attempts have been made in the past to determine $m_s$ in various frameworks, e.g. QCD sum rules \cite{OLD1}-\cite{OLD2}, and Lattice QCD \cite{LATTICE}. The most recent QCD sum rule determinations from the pseudoscalar channel  have made use of state of the art results in perturbative QCD (PQCD) to five-loop order \cite{PQCD5}. In spite of this, the real uncertainty in the value of $m_s$ remains high due to the hadronic resonance sector. In fact, beyond the kaon pole, the pseudoscalar hadronic spectral function is not known from direct experimental mesurements. Two radial excitations of the kaon have been observed \cite{PDG}, and a certain amount of theoretical input has gone into attempts to build a reasonable spectral function incorporating these resonances. However, inelasticity and non-resonant background are realistically impossible to model. This constitutes a form of systematic uncertainty seriously limiting the precision of these determinations. In summary, current information on the QCD side of the sum rules is not matched in quality by the pseudoscalar hadronic sector. An attempt to rectify this situation has been made recently \cite{COND} in the form of a new QCD Finite Energy Sum Rule (FESR) involving as integration kernel a second degree polynomial which is required to vanish at the peaks of the two pseudoscalar resonances. As a result of this, the kaon pole and the QCD contributions dominate the FESR; the importance of the hadronic resonance sector being reduced by up to an order of magnitude.
This FESR was used in \cite{COND} to determine the scalar and pseudoscalar correlators at zero momentum, and the strange quark condensate. An upper bound on the strange quark mass was also obtained there, e.g. for the running mass at a scale of $ 2\; \mbox{GeV}$ this bound is

%Eq.1
\begin{eqnarray}
m_{s} (2 \; \mbox{GeV}) \;  \leq \;\Bigg\{ 
\begin{array}{lcl}
121  \; \mbox{MeV} \; \; \; (\Lambda_{QCD} = 330 \; \mbox{MeV})\\[.3cm]
105  \; \mbox{MeV} \; \; \; (\Lambda_{QCD} = 420 \; \mbox{MeV}) \;.\end{array}
\end{eqnarray}

In this paper we use this FESR in the pseudoscalar channel to determine the value of strange quark mass to five-loop order in PQCD, and including the leading vacuum condensates. We use the framework of Fixed Order Perturbation Theory (FOPT), as well as Contour Improved Perturbation Theory (CIPT). 
We find that as a result of the integration kernel in the FESR,  the hadronic resonance contribution is considerably reduced relative to the kaon pole. The latter is of the same order as the PQCD contribution, and both are up to a factor five bigger than the resonance term, this being comparable to the gluon condensate term. Numerically, the resonances add roughly 10\% to the quark mass, relative to the value obtained from the kaon pole and PQCD, while the gluon condensate subtracts a similar amount, and the light-quark condensate reduces it by another 1-2 \%. Higher dimensional condensates and higher order quark-mass terms contribute negligible amounts. Results from FOPT for the running strange quark mass at a fixed scale are reasonably stable in a wide range of values of $s_0$, the upper limit of integration in the FESR ($s_0 \simeq 2.5 - 4.5 \; \mbox{GeV}^2$). However, in CIPT the stability is remarkable in an even wider range, e.g. $m_s$(2 GeV) changes by less than 1\% in the range $s_0 \simeq 1.0 - 4.0 \; \mbox{GeV}^2$. 
There is a very strong correlation between $m_s$ and the value of the QCD scale $\Lambda$, which produces most of the uncertainty in the result for the strange quark mass (roughly 16 \%). However, unlike the situation in the hadronic resonance sector, the uncertainty in $\Lambda$ can be reduced, in principle, by improving the accuracy of the theoretical input, as well as of the data used in its determination. 
 
\section{Fixed Order Perturbation Theory}
We first introduce the correlator of  axial-vector divergences 

%Eq.2
\begin{equation}
\psi_{5} (q^{2})   = i \, \int\; d^{4}  x \; e^{i q x} \; 
<|T(\partial^\mu A_{\mu}(x) \;, \; \partial^\nu A_{\nu}^{\dagger}(0))|> \;,
\end{equation}

where $\partial^\mu A_{\mu}(x) = (m_s + m_u) :\overline{s}(x) \,i \, \gamma_{5}\, u(x):\;$ is the divergence of the  axial-vector current. To simplify the notation  we shall use in the sequel $m_s + m_u \equiv m$. Finite Energy Sum Rules (FESR) involving this correlator follow from Cauchy's theorem in the complex energy-squared, s - plane (see Fig. 1), i.e.

%Eq.3
\begin{eqnarray}
0 &=&
\int_{0}^{s_0}
ds \frac{1}{\pi} Im \;\psi_{5}(s)+ \frac{1}{2\pi i}
\oint_{C(|s_0|)}
ds \;\psi_{5}(s) \nonumber \\ [.3cm]
&\simeq&
\int_{0}^{s_0}
ds \frac{1}{\pi} Im \;\psi_{5}^{.}(s)+\frac{1}{2\pi i}
\oint_{C(|s_0|)}
ds \;\psi_{5}^{QCD}(s) \, ,\label{CAUCHY1}
\end{eqnarray}

and  the contour integral is performed over a large circle where the exact $\psi_{5}(s)$ can  be safely replaced by its QCD counterpart $\psi_{5}^{QCD}(s)$. We introduce now an integration kernel in the form of a second degree polynomial

%eq.4
\begin{equation}
\Delta_5(s) = 1 - a_0 \;s - a_1\; s^2 \;,
\end{equation}

where $a_{0}$, and $a_1$ are free parameters to be fixed by the requirement that $\Delta_5(M_1^2) = \Delta_5(M_2^2) = 0$, with $M_{1,2}$ the masses of the two resonances in the strange pseudoscalar channel, K(1460) and K(1830) \cite{PDG}. This gives 

%eq.5
\begin{equation}
a_0 = 0.768  \; \mbox{GeV}^{-2} \; \;  a_1 = - 0.140 \;\; \mbox{GeV}^{-4} \;.
\end{equation}

\begin{figure}
[ht]
\begin{center}
\includegraphics[height=2.5in, width=2.5in]
{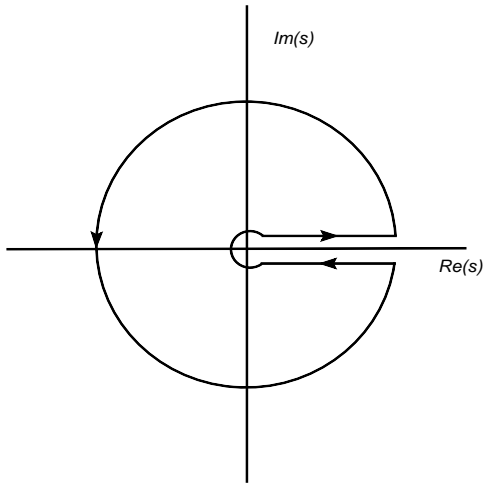}
\caption{Integration contour in the complex s-plane.}
\end{center}
\end{figure}

Writing explicitly the kaon pole, Cauchy's theorem, Eq. 3, becomes

%Eq.6
\begin{eqnarray}
- \frac{1}{2\pi i}
\oint_{C(|s_0|)}
ds \;\psi_{5}^{QCD}(s)\; \Delta_5(s) &=& 2\; f_K^2 \; M_K^4\; \Delta_5(M_K^2)  \nonumber \\ [.3cm]
&+&
\int_{s_{th}}^{s_0}
ds \;\frac{1}{\pi} \;Im \;\psi_{5}(s)|_{RES}\;\Delta_5(s) \, , 
\end{eqnarray}

where $s_{th} = (M_K + M_\pi)^2$ is the resonance threshold, and $f_K = (1.22 \pm 0.01) f_\pi = 113 \pm 1 \;\mbox{MeV}$. The introduction of the integration kernel $\Delta_5(s)$ is expected to reduce the importance of the hadronic resonances in the determination of $m_s$. A {\it posteriori}, this is in fact the case: in the stability region for $m_s(s_0)$ the resonance contribution is roughly a factor five smaller than the pole term, as well as  the PQCD piece.\\

In the framework of FOPT $\alpha_s$ and $m_s$ are taken as constants, and only terms involving powers of $log(-s/\mu^2)$
contribute to the contour integral. The renormalization group summation of leading logs is only carried out after the contour integration by setting $\mu^2 = - s_0$.  The PQCD result for $\psi_5(s)$ up to four-loop order  has been known for quite some time \cite{PQCD4}, while the five-loop expression for its second derivative has been obtained recently \cite{PQCD5}. Integrating the latter twice provides $\psi_5(s)$ up to a non-contributing polynomial. We define

%eq.7
\begin{equation}
\delta_5(s_0)|_{QCD} \equiv \, - \frac{1}{2 \pi i} \oint_{C(|s_0|)} ds\; \Delta_5(s)\; \psi_5(s)|_{QCD} \;,
\end{equation}
 
and obtain in PQCD the following results  

%eq.8
\begin{equation}
\delta_5(s_0)|_{1LOOP} =  \frac{\overline{m}^2(s_0)}{16 \pi^2} \; C_{01}\;\left[\frac{s_0^2}{2} - a_0\, \frac{s_0^3}{3} - a_1\, \frac{s_0^4}{4} \right] \:,
\end{equation}

%eq.9
\begin{eqnarray}
\delta_5(s_0)|_{2LOOP} &=&  \frac{\overline{m}^2(s_0)}{16 \pi^2} \; \frac{\alpha_s(s_0)}{\pi}\; \left[C_{11}\,\Big(\frac{s_0^2}{2} - a_0\, \frac{s_0^3}{3} - \; a_1\; \frac{s_0^4}{4}\Big) \right.\nonumber \\ [.3cm]
 &-& \left.  2 \;C_{12} \;\Big(\frac{s_0^2}{4} - a_0\, \frac{s_0^3}{9} - \; a_1\; \frac{s_0^4}{16}\Big) \right] \:,
\end{eqnarray}

%eq.10
\begin{eqnarray}
\delta_5(s_0)|_{3LOOP} &=&  \frac{\overline{m}^2(s_0)}{16 \pi^2} \; [\frac{\alpha_s(s_0)}{\pi}]^2 \; \left\{ C_{21} \left(\frac{s_0^2}{2} - a_0\, \frac{s_0^3}{3} - \; a_1\; \frac{s_0^4}{4}\right)   \right. \nonumber \\ [.3cm]
&-& \left. 2 \;C_{22} \;\left(\frac{s_0^2}{4} - a_0\, \frac{s_0^3}{9} - \; a_1\; \frac{s_0^4}{16}\right) -6\; C_{23} \left[ \frac{s_0^2}{2} \left( \frac{\pi^2}{6} - \frac{1}{4}\right)
 \right. \right. \nonumber \\ [.3cm]
 &-& \left. \left.    a_0 \;\frac{s_0^3}{3}\; \left(\frac{\pi^2}{6} - \frac{1}{9}\right) 
 -  a_1 \;\frac{s_0^4}{4}\;\left( \frac{\pi^2}{6} - \frac{1}{16}\right) \right]   \right\} \:,
\end{eqnarray}

%eq.11
\begin{eqnarray}
&\delta_5(s_0)|_{4LOOP}& =  \frac{\overline{m}^2(s_0)}{16 \pi^2}\;  [\frac{\alpha_s(s_0)}{\pi}]^3 \; \left\{C_{31} \left(\frac{s_0^2}{2} - a_0\, \frac{s_0^3}{3} - \; a_1\; \frac{s_0^4}{4}\right)   \right. \nonumber \\ [.3cm]
&-& \left. 2 \;C_{32} \;\left(\frac{s_0^2}{4} - a_0\, \frac{s_0^3}{9} - \; a_1\; \frac{s_0^4}{16}\right) - 6 \;C_{33} \left[ \frac{s_0^2}{2} \left( \frac{\pi^2}{6} - \frac{1}{4}\right)
 \right.  \right. \nonumber \\ [.3cm]
&-& \left. \left.   a_0 \;\frac{s_0^3}{3} \left(\frac{\pi^2}{6} - \frac{1}{9}\right)- a_1\; \frac{s_0^4}{4} \left(\frac{\pi^2}{6} - \frac{1}{16}\right)\right]
+ 24\; C_{34} \left[ \frac{s_0^2}{2} \right. \right. \nonumber \\ [.3cm] 
 &\times& \!\!\!\!\!\!\left.\left.\left(\frac{\pi^2}{6} - \frac{1}{4}\right) - a_0 \frac{s_0^3}{9} \left(\frac{\pi^2}{6} - \frac{1}{9}\right)- a_1 \frac{s_0^4}{16} \left(\frac{\pi^2}{6} - \frac{1}{16}\right)\right] \right\}.
\end{eqnarray}

%eq.12
\begin{eqnarray}
\delta_5(s_0)|_{5LOOP} &=&  \frac{\overline{m}^2(s_0)}{16 \pi^2}\;  [\frac{\alpha_s(s_0)}{\pi}]^4 \; \left\{C_{41} \left(\frac{s_0^2}{2} - a_0\, \frac{s_0^3}{3} - \; a_1\; \frac{s_0^4}{4}\right) \right.  \nonumber \\ [.3cm]
&-& \left. 2 \;C_{42} \;\left(\frac{s_0^2}{4} - a_0\, \frac{s_0^3}{9} - \; a_1\; \frac{s_0^4}{16}\right) - 6 \;C_{43} \left[ \frac{s_0^2}{2} \left( \frac{\pi^2}{6} - \frac{1}{4}\right) \right.\right.  \nonumber \\ [.3cm]
&-& \left. \left.   a_0 \;\frac{s_0^3}{3} \left(\frac{\pi^2}{6} - \frac{1}{9}\right)- a_1\; \frac{s_0^4}{4} \left(\frac{\pi^2}{6} - \frac{1}{16}\right)\right]
+ 24\; C_{44} \left[ \frac{s_0^2}{4} \right. \right. \nonumber \\ [.3cm]
&\times& \left. \left. \left( \frac{\pi^2}{6} - \frac{1}{4}\right) - a_0 \frac{s_0^3}{9} \left(\frac{\pi^2}{6} - \frac{1}{9}\right)- a_1 \frac{s_0^4}{16} \left(\frac{\pi^2}{6} - \frac{1}{16}\right)\right] \right. \nonumber \\ [.3cm]
 &+& \left. 120 \; C_{45} \left[ \frac{s_0^2}{2} \left( \frac{\pi^4}{120} - \frac{\pi^2}{24} + \frac{1}{16} \right)
 - a_0\; \frac{s_0^3}{3} \left( \frac{\pi^4}{120} \right. \right. \right. \nonumber \\ [.3cm]
&-& \left.\left.\left.\frac{\pi^2}{54} + \frac{1}{81} \right)
- a_1\; \frac{s_0^4}{4} \left( \frac{\pi^4}{120} - \frac{\pi^2}{96} - \frac{1}{256} \right) \right] \right\} \;,
\end{eqnarray}

where $m \equiv m_s + m_u$, and the constants $C_{ij}$ above, for three quark flavours, are: $C_{01} = 6$, $C_{11} = 34$, $C_{12} = - 6$, $C_{21} = - 105 \;\zeta(3) + 9631/24$, 
$C_{22} = - 95$, $C_{23} = 17/2$, $C_{31}= 4748953/864 - \pi^4/6 - 91519\; \zeta(3)/36 + 715 \;\zeta(5)/2$,
$C_{32} = - 6 \;[4781/18 - 475 \;\zeta(3)/8]$, $C_{33} = 229$, $C_{34} = - 221/16$, $C_{41} = 33 532.26$, $C_{42} = - 15 230. 6451$, $C_{43} = 3962.45493$, $C_{44} = - 534.052083$, $C_{45} = 24.1718750$, and $\zeta(x)$ is Riemann's zeta function. Regarding the value of $\Lambda$ entering $\alpha_s(s_0)$, since we are dealing with three quark flavours, it is simpler to determine $\Lambda_{QCD}$ from the strong coupling obtained from $\tau$-decay \cite{PDG}, \cite{DAVIER}: $\alpha_s(M_\tau^2) = 0.31 - 0.36$, which gives $\Lambda_{QCD} = 330 - 420\; \mbox{MeV}$.\\

The leading non-perturbative contributions are due to the gluon and the light-quark condensates, which give

%Eq.13
\begin{equation}
\delta_5(s_0)|_{<G^2>} = \,\frac{\overline{m}^2(s_0)}{8}\;
\langle \frac{\alpha_s}{\pi} G^2\rangle \left[ 1 + \frac{\alpha_s(s_0)}{\pi} \left( \frac{11}{2} + 2 \,a_0\, s_0 + a_1\, s_0^2 \right) \right]\;,
\end{equation}

%Eq.14
\begin{equation}
\delta_5(s_0)|_{<\bar{u}u>} = \overline{m}^2(s_0)\,
\langle m_s \bar{u}u\rangle \left[ 1 + \frac{\alpha_s(s_0)}{\pi} \left( \frac{14}{3} + 2 \,a_0\, s_0 + a_1\, s_0^2 \right) \right]\;,
\end{equation}

where $\langle \frac{\alpha_s}{\pi} G^2 \rangle \simeq 0.06\; \mbox{GeV}^4$, and $\langle \bar{q}\,q \rangle \simeq ( - 250 \;\mbox{MeV})^3$ \cite{CADTAU}.
We find that terms of $\cal{O}$$(m^4)$ and higher, as well as the strange quark condensate, give negligible contributions in the region of stability (which turns out to be $ s_0 \simeq 2.5 - 4.5 \; \mbox{GeV}^2$). This is also the case for the condensates of dimension-six and higher.\\

Turning to the hadronic sector, the spectral function in the pseudoscalar channel involves in addition to the kaon pole, at least two radial excitations, the K(1460) and K(1830) both with widths of about 250 MeV \cite{PDG}. We follow the procedure outlined in \cite{CAD1}, where the resonance part of the spectral function is written as a linear combination of two Breit-Wigner forms normalized at threshold according to chiral perturbation theory. The latter incorporates the resonant sub-channel $K^*(892)-\pi$ which is important due to the narrow width of the $K^*(892)$. Other embellishments are certainly possible, but then the presence of the integration kernel $\Delta_5(s)$ in Eq.(6) makes these attempts unnecessary. In fact, the resonance contribution to Eq.(6) in a wide range of values of $s_0$ is up to a factor five smaller than the PQCD term, and a similar factor smaller than the kaon pole contribution.\\

\begin{figure}
[ht]
\begin{center}
\includegraphics[width=\columnwidth]{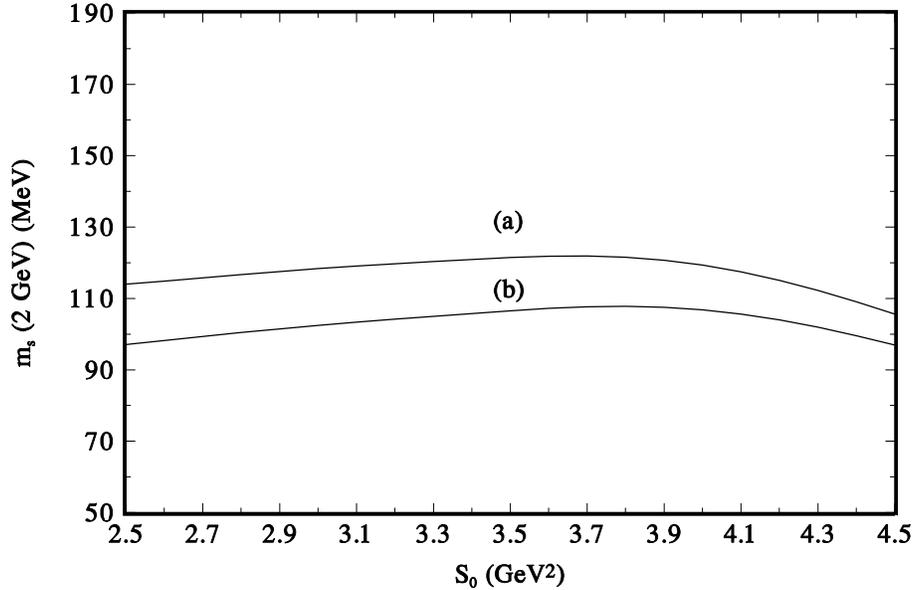}
\caption{Running strange quark mass in FOPT at a scale of 2 GeV as a function of $s_0$. Curve (a) is for $\Lambda = 330$ MeV, and curve (b) for $\Lambda = 420$ MeV.}
\end{center}
\end{figure}

Adding up Eqs.(8)-(14) gives the left hand side of the FESR, Eq.(6), which leads to the results for $m_s$(2 GeV) shown in Fig. 2. Curve (a) corresponds to $\Lambda = 330$ MeV, and curve (b) to $\Lambda = 420$ MeV. In obtaining the results shown in Fig. 2, the light quark mass has already been subtracted; its value can be safely estimated from the chiral perturbation theory ratio \cite{RATIO}: $\frac{m_s}{m_q} = 24.4 \pm 1.5$, where $m_q = (m_u + m_d)/2$.
For the running coupling needed in Eqs.(9)-(14), enough precision is obtained using the four-loop result. We have also achieved enough precision with the four-loop result to convert the running quark mass $\overline{m_s}(s_0)$ into $m_s$(2 GeV). Numerically, the latter is $m_s$(2 GeV) $ = 114 \pm 8$ MeV ($102 \pm 6$ MeV), for $\Lambda = 330$ MeV ($\Lambda = 420$ MeV), respectively. The breakdown of the various factors contributing to the quark mass is as follows. To the basic result from PQCD and the kaon pole, the hadronic resonances add roughly 10 \% to the mass, the gluon condensate reduces the mass by a similar amount, and the light quark condensate by an additional 1-2 \%. The errors quoted above are only due to variations of $m_s$(2 GeV) inside the stability region $s_0 = 2.5 - 4.5 \; \mbox{GeV}^2$. A more realistic error can be established by assuming that the resonance contribution to the quark mass has been understimated or overestimated by a certain amount. Choosing this amount conservatively as 30 \% produces an additional uncertainty in $m_s$ (2 GeV) of $\pm 4$ MeV. In addition, assuming that the unknown six-loop contribution is comparable to the five-loop one would reduce $m_s$ (2 GeV) by roughly 2 MeV. Combining all uncertainties and using the breakdown described above, the final results for the quark mass are

%Eq.15
\begin{eqnarray}
m_{s} (2 \; \mbox{GeV}) \;  = \;\Bigg\{ 
\begin{array}{lcl}
114 \pm 14  \; \mbox{MeV} \; \; \; (\Lambda_{QCD} = 330 \; \mbox{MeV})\\[.3cm]
102 \pm 11 \; \mbox{MeV} \; \; \; (\Lambda_{QCD} = 420 \; \mbox{MeV}) \;.\end{array}
\end{eqnarray}

\section{Contour Improved Perturbation Theory}

Contour Improved Perturbation Theory \cite{CIPT} has been shown to provide better convergence than FOPT in the QCD analysis of the vector and axial-vector correlators in tau-lepton decay. We find this to be also the case for our analysis of the contour integral in Eq.(3). Unlike the case of FOPT, where $\alpha_s(s_0)$ is frozen in Cauchy's contour integral and the Renormalization Group (RG) is implemented after integration, in CIPT $\alpha_s$ is running and the RG is used before integrating. This is done  through a single-step numerical contour integration and using as input the strong coupling
obtained  by solving numerically the Renormalization Group Equation  for $\alpha_s(-s)$ . This technique achieves a partial resummation of the higher order logarithmic integrals, and improves the convergence of the PQCD series. CIPT has been used successfully in QCD analyses of tau-lepton hadronic decays \cite{CADTAU}, \cite{CIPT}. In the case of the pseudoscalar correlator involving the running quark mass as an overall multiplicative factor, implementation of CIPT requires that not only the running coupling but also the running quark mass be integrated around the Cauchy contour. The running quark mass can be computed at each step by solving numerically the  corresponding RGE.
To establish  notation and conventions, we write the RGE for the coupling as

%Eq.16
\begin{equation}
s \; \frac{d \, a_s(-s)}{d s} = \beta (a_s) = - \sum_{N=0} \beta_N \; a_s(-s)^{N+2} \;,
\end{equation}

where $a_s \equiv \alpha_s/\pi$, and for three quark flavours $\beta_0 = 9/4$, $\beta_1 = 4$, $\beta_2 = 3863/384$,
$\beta_3 = (421797/54 + 3560 \zeta(3))/256$. In the complex s-plane $s = s_0\, e^{ix}$ with the angle $x$ defined in the interval $x \in (- \pi, \pi)$. The RGE then becomes

%Eq.17
\begin{equation}
\frac{d \, a_s(x)}{d x} = - i \sum_{N=0} \beta_N \; a_s(x)^{N+2} \;,
\end{equation}

This RGE can be solved numerically using e.g. a modified Euler method, providing as input $ a_s (x=0) = a_s (- s_0)$. Next, the RGE for the quark mass is given by

%Eq.18
\begin{equation}
\frac{s}{m} \; \frac{d \, m(-s)}{d s} = \gamma (a_s) = - \sum_{M=0} \gamma_M \; a_s^{M+1} \;,
\end{equation}

where for three quark flavours $\gamma_0 = 1$, $\gamma_1 = 182/48$, $\gamma_2 = [8885/9 - 160 \,\zeta(3)]/64$, $\gamma_3 = [2977517/162 - 148720 \,\zeta(3)/27 + 2160 \,\zeta(4) - 8000\, \zeta(5)/3]/256$. With the aid of Eqs. (16)-(17) the above equation can be converted into a differential equation for $m(x)$ and integrated, with the result

%Eq.19
\begin{equation}
m(x) = m(0) \;exp \Big\{ - i \int_0^x dx' \sum_{M=0} \gamma_M \, [a_s(x')]^{M+1}\Big\}\;,
\end{equation}

where the integration constant $m(0)$ is identified as the overall multiplicative quark mass in the expression for the pseudoscalar correlator, i.e. $m \equiv [m_s(s_0) + m_u(s_0)]$. Cauchy's theorem, and the resulting FESR will be written for the second derivative of $\psi_5(s)$, in which case it is straightforward to show the following identity

%Eq.20
\begin{equation}
\oint ds \, g(s) \, \psi_5(s) = \oint ds \, [F(s) - F(s_0)] \;\psi_5''(s) \;,
\end{equation}

where

%Eq.21
\begin{equation}
F(s) = \int _0^s ds' \left[ \int_0^{s'} ds'' g(s'') - \int_0^{s_0} ds'' g(s'')\right] \;,
\end{equation}

and $g(s)$ is an arbitrary analytic function which we choose as $g(s) = \Delta_5(s)$, with $\Delta_5(s)$ given in Eq.(4). In this case instead of Eq.(6) the FESR becomes

%Eq.22
\begin{eqnarray}
- \frac{1}{2\pi i}
\oint_{C(|s_0|)}
&ds& \psi_{5}^{'' QCD}(s)\,[F(s) - F(s_0)] = 2\; f_K^2 \; M_K^4\; \Delta_5(M_K^2)  \nonumber \\ [.3cm]
&+&
\frac{1}{\pi} \; \int_{s_{th}}^{s_0}
ds \; Im \;\psi_{5}(s)|_{RES}\;\Delta_5(s) \, , 
\end{eqnarray}

where

%Eq.23
\begin{equation}
F(s) = - s \left(s_0 - a_0\,\frac{s_0^2}{2} - a_1\, \frac{s_0^3}{3} \right) + \frac{s^2}{2} - a_0\, \frac{s^3}{6} - a_1\, \frac{s^4}{12} \;,
\end{equation}

%Eq.24
\begin{equation}
F(s_0) = - \frac{s_0^2}{2} +  a_0\, \frac{s_0^3}{3} + a_1\, \frac{s_0^4}{4} \;,
\end{equation} 

%Eq.25
\begin{eqnarray}
\psi_5^{'' PQCD}(Q^2) &=& \frac{3}{8 \pi^2}\; \frac {\overline{m}^2(Q^2)}{ Q^2}\;  \left\{ 1 +\; \frac{11}{3}\; \frac{\alpha_s(Q^2)}{\pi}\;  +\; (\frac{\alpha_s(Q^2)}{\pi})^2 \;\left[ -  \frac{35}{2}\; \zeta(3)
\right.\right. \nonumber \\ [.3cm]
&+& \left. \left. \frac{5071}{144}  \right] + O (\alpha_s^3) \right\} \;,
\end{eqnarray}

with $Q^2 \equiv - q^2$, and Renormalization Group improvement has been used to dispose of the logarithmic terms.
The rather long four- and five-loop expressions can be found in \cite{PQCD5} and \cite{PQCD4}. The left hand side of Eq.(22) in PQCD can be written as

%Eq.26
\begin{eqnarray}
\delta_5(s_0)|_{PQCD} &\equiv& - \frac{1}{2\pi i}
\oint_{C(|s_0|)}
ds \;\psi_{5}^{'' PQCD}(s)\,[F(s) - F(s_0)]   \nonumber \\ [.3cm]
&=& \frac{\overline{m}^2(s_0)}{16 \pi^2} \,\sum_{j=0}^4 K_j \;
\frac{1}{2 \pi} \;
  \int_{-\pi}^{\pi} dx \;  \Big[ F(x) - F(s_0) \Big]  \nonumber \\ [.3cm]
&\times&  [a_s(x)]^j \;
exp \Bigg[ - 2 i \sum_{M=0} \gamma_M \int_0^x\; dx' \;[a_s(x')]^{M+1} \Bigg]  \;,
\end{eqnarray}

where $K_0= C_{01}$, $K_1 = C_{11} + 2 C_{12}$, $K_2 = C_{21} + 2 C_{22}$, $K_3 = C_{31} + 2 C_{32}$, $K_4 = C_{41} + 2 C_{42}$, with $C_{ik}$ defined after Eq.(12), and 

%Eq.27
\begin{equation}
F(x) = \sum_{N=1}^4 (-)^N \; b_N \; s_0^N \; e^{iNx} \;,
\end{equation}

and $b_1= -(s_0 - a_0 s_0^2/2 - a_1 s_0^3/3)$, $b_2 = 1/2$, $b_3 = - a_0/6$, and $b_4 = - a_1/12$. The contribution of the gluon condensate to the left hand side of Eq.(22) is

%Eq.28
\begin{eqnarray}
\delta_5(s_0)|_{<G^2>} &=& \frac{1}{4} \;\;\frac{\overline{m}^2(s_0)}{s_0^2} \; \;\langle \frac{\alpha_s}{\pi} G^2\rangle|_{\mu_0}\;\;
\frac{1}{2 \pi} \;\int_{-\pi}^\pi dx \;e^{-2 i x} \nonumber \\ [.3cm]
&\times& \Big[F(x) - F(s_0)\Big] \Big[ 1 + \frac{16}{9} a_s(\mu_0) + \frac{121}{18} a_s(x)\Big] \nonumber \\ [.3cm]
&\times& exp  \Bigg[ - 2 i \sum_{M=0} \gamma_M  \int_0^x dx' [a_s(x')]^{M+1} \Bigg] \;,
\end{eqnarray}

where the scale $\mu_0 \simeq 1 \; \mbox{GeV}^2$ appears in connection with the removal of logarithmic quark mass singularities (see \cite{OLD2}). The light-quark condensate contribution is given by

%Eq.29
\begin{eqnarray}
\delta_5(s_0)|_{<\bar{q} q>} &=& - 2 \;\;\frac{\overline{m}^2(s_0)}{s_0^2} \; \;\langle m_s \overline{q} q \rangle|_{\mu_0}\;\;
\frac{1}{2 \pi} \;\int_{-\pi}^\pi dx \;e^{-2 i x} \nonumber \\ [.3cm]
&\times& \Big[F(x) - F(s_0)\Big] \Big[ 1 + \frac{23}{3} a_s(x)\Big] \nonumber \\ [.3cm]
&\times& exp  \Bigg[ - 2 i \sum_{M=0} \gamma_{M}  \int_0^x dx' [a_s(x')]^{M+1} \Bigg] \;.
\end{eqnarray}

\begin{figure}
[ht]
\begin{center}
\includegraphics[width=\columnwidth]{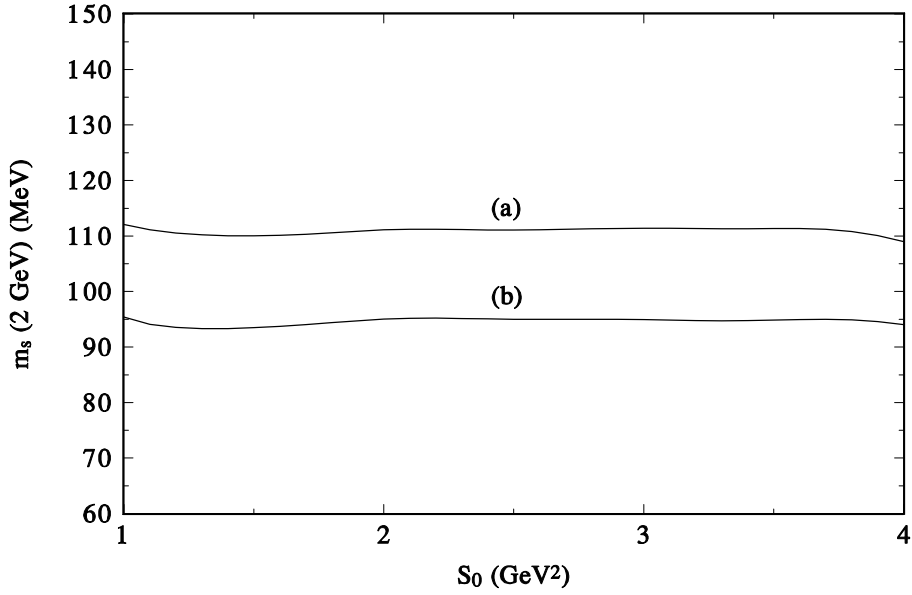}
\caption{Running strange quark mass in CIPT at a scale of 2 GeV as a function of $s_0$. Curve (a) is for $\Lambda = 330$ MeV, and curve (b) for $\Lambda = 420$ MeV.}
\end{center}
\end{figure}

Substituting Eqs. (26), (28) and (29) in the FESR, Eq.(22), completes the expression giving the running quark mass  $\overline{m}(s_0)$. The breakdown of the various contributions is similar to the case of FOPT. The Kaon pole and PQCD are the leading terms in the FESR, the hadronic resonances add 8 - 9 \% to the mass, the gluon condensate reduces it by roughly the same amount, and the light-quark condensate induces a further reduction of 1 -2 MeV. As expected, the convergence of the PQCD series is much better than the FOPT counterpart. The results are shown in Fig. 3  for the running  strange quark mass at 2 GeV. Curves (a) and (b) correspond to $\Lambda = 330 \;(420)\; \mbox{MeV}$, respectively. The stability of the results, and the wide range of the stability region are quite remarkable. In fact, $m_s$ (2 GeV) changes by not more than 1-2 MeV in the interval $s_0 = 1 - 4 \; \mbox{GeV}^2$. This stability is present at each loop level in PQCD, i.e. the quark mass as a function of $s_0$ is essentially flat if computed at one loop level, one- plus two-loop level, etc.. 

To arrive at a reasonable error, we proceed as before in FOPT, and assume that the  hadronic resonance contributions could be underestimated or overestimated  by up to 30 \%. In this case, the uncertainty in $m_s$ (2 GeV) for $\Lambda =$ 330 (420) MeV, would be $ \pm 3 \;(2)$ MeV. In addition, if one were to assume the extreme scenario in which the six-loop contribution would be comparable to the five-loop one, then a further reduction in the mass by roughly 3 MeV would result for both values of $\Lambda$. Combining these two uncertainties linearly, the final results are

%Eq.30
\begin{eqnarray}
m_{s} (2 \; \mbox{GeV}) \;  = \;\Bigg\{ 
\begin{array}{lcl}
111 \pm 6 \; \mbox{MeV} \; \; \; (\Lambda_{QCD} = 330 \; \mbox{MeV})\\[.3cm]
95 \pm 5 \; \mbox{MeV} \; \; \; \;\;(\Lambda_{QCD} = 420 \; \mbox{MeV}) \;.\end{array}
\end{eqnarray}

A breakdown of the contribution of each loop-order in PQCD to the quark mass $m_s$ (2 GeV) is as follows. For $\Lambda = 330$ MeV, the pole plus resonances plus condensates and only the one-loop contribution would give
$m_s$ (2 GeV) = 152 MeV, including  the two-loop term reduces this to $m_s$ (2 GeV) = 129 MeV, with the three-loop it gives $m_s$ (2 GeV) = 120 MeV, with up to four loops this becomes
$m_s$ (2 GeV) = 115 MeV, and finally with all five loops  the result is $m_s$ (2 GeV) = 111 MeV. Similar relative differences are found for $\Lambda = 420$ MeV. As seen from Eq. (30) the main uncertainty is due to the value of $\Lambda$. Finally, we have considered the impact on the above results from uncertainties in the vacuum condensates as well as inclusion of higher order quark mass corrections.  Allowing the extreme case of an uncertainty of a factor two in the gluon condensate produces changes in $m_s$(2 GeV) well within the errors given in Eq.(30). The contribution of the strange quark condensate is negligible, and that of the light (up- or down-) quark condensate, which is known accurately, is at the level of 1\%. Higher order quark mass corrections, as well as vacuum condensates of dimension $d=6$, have no impact on the results and can be safely ignored in this method. The results above satisfy comfortably the upper bounds given in Eq.(1). In this particular application, CIPT has proven to be far better than FOPT. Although the results from both methods agree within errors, CIPT leads to a remarkable stability of $m_s$(2 GeV), in a remarkable wide range of values of $s_0$. For this reason, we would not advocate combining the results from both methods.

\section{Conclusions}

The main advantage of using  pseudoscalar correlators to determine the quark masses is that they enter the QCD expressions as overall multiplicative factors, rather than as corrections to a leading term. In addition, they involve the pseudoscalar pole with parameters well known from experiment. Unfortunately, there is no direct experimental information on the  hadronic resonance spectral function, except for the masses and widths of the first few resonances. This information is not enough to reconstruct reliably these spectral functions, as inelasticity and non-resonant background effects are realistically impossible to guess. For this reason, quark mass determinations from pseudoscalar correlators are affected by endemic systematic uncertainties not subject to improvement. In this paper we have used a new QCD FESR \cite{COND} for the strange pseudoscalar correlator, involving as integration kernel a second degree polynomial which is required to vanish at the peaks of the first two radial excitations of the kaon. As a result of this, the relative importance of the hadronic resonance sector in the determination of the strange quark mass is considerably reduced. In fact, this contribution turns out to be up to a factor five smaller than the leading contributions from PQCD and from the well known kaon pole. We have used the techniques of FOPT as well as CIPT to compute the integrals in the complex energy (squared) plane. The latter method is far superior to the former, and gives a running mass at a fixed scale which is remarkably stable in a very wide range of $s_0$, the radius of the integration contour in the complex plane. For instance, for the strange quark mass at 2 GeV we find 
$m_s$ (2 GeV) = 111 (95) MeV, for $\Lambda$ = 330 (420) MeV, respectively, in the wide range $s_0 = 1 - 4 \;\mbox{GeV}^2$.
To  arrive at a reasonable, but still conservative estimate of the uncertainties we have assumed that (a) the resonance parametrization might be an underestimate or an overestimate  of the hadronic spectral function of up to 30 \%, and (b) the unknown six-loop PQCD term could be comparable to the five-loop term. Each of these assumptions induces an uncertainty in the quark mass at the level of 3 \%. Adding them linearly gives the results in Eq. (30). The main uncertainty is then due to $\Lambda$. However, unlike the hadronic resonance spectral function, this error is subject to improvement. The results from FOPT, Eq. (15), are expectedly in agreement with those from CIPT. However, in FOPT an additional sizable uncertainty arises from the variation of $m_s$ (2 GeV) in the (narrower) stability range.
Results from both methods satisfy the upper bound Eq. (1).
Comparison of our results with previous determinations is made somewhat difficult due to various reasons. Some of the very old determinations were afflicted by logarithmic quark mass singularities in the correlators. This issue was only clarified in \cite{OLD2}. In addition, the values of $\Lambda$ used in the past were much lower than at present. Given the strong correlation between $\Lambda$ and the quark mass, this becomes a serious issue. Next, knowledge of the PQCD contribution has improved considerably over the years, from two-loop level to the current five-loop level. Last, but not least, the systematic uncertainties due to a lack of direct experimental information on the hadronic resonance spectral function may have been underestimated in the past. In any case, comparing the results in Eq. (30) with the most recent determinations \cite{OLD1}, \cite{LATTICE}, \cite{PQCD5}, shows very good overall agreement.\\

In closing we wish to mention that  in some applications of FESR, e.g. in tau-decay, perturbative QCD does not appear to hold close to the real axis. It is not entirely clear whether there is a problem with the data, or with PQCD itself. In any case this has led to the proposal of weighted FESR with weight functions vanishing at $s = s_0$ \cite{MALT}. These so called $\it{pinched}$ FESR improve considerably the saturation of the Weinberg sum rules, and resolve some inconsistencies in the determination of vacuum condensates in the vector and axial-vector channels. \\

With this background it is reasonable to investigate the impact of such kernels in the determination of the strange quark mass discussed here. First of all, the rate at which a hadronic spectral function approaches its PQCD limit is channel dependent. Presumably, potential duality violations share this feature. In the framework of the method discussed here, it turns out that the ratio of the QCD and the hadronic contributions, which equals the square of the running quark mass, leads to a value for $m_s$(2 GeV) which is remarkably stable as a function of the upper limit of integration $s_0$, in an also remarkable wide range $s_0 \simeq 1 - 4\; \mbox{GeV}^2$. Such quality is hardly found in typical QCD sum rule applications. It is then reasonable to conclude that in this particular channel, and using our integration kernel, duality appears to be well satisfied. It should be stressed that the motivation for introducing the kernel Eq.(4)in the pseudoscalar channel is rather different from that for the $\it{pinched}$ kernel in tau-decays. In fact, the hadronic spectral  functions in the latter case are known from experiment, while this is not the case for the pseudoscalar channel.\\

In any case, and to continue looking at this issue, we may  consider the direct product of the kernel Eq.(4) and a $\it{pinched}$ one. The first undesirable result of such a procedure is that the kaon pole contribution now becomes a function of $s_0$, and is  numerically reduced. This does not happen to the Weinberg sum rules, as they are  valid in the chiral limit, in which case the pseudoscalar meson pole contribution to the spectral function involves the delta function $\delta(s)$. As a result of this, the $\it{pinched}$ kernel does not affect this contribution. This behaviour of the kaon pole contribution is contrary to the spirit of the method used here. In fact, since there is accurate experimental information on this pole, one wishes to enhance its contribution rather than reduce it.
In any case, using this additional kernel we have studied the convergence of the PQCD series, and the relative contributions of the various terms, e.g. vacuum condensates, higher order quark-mass corrections, pseudoscalar meson pole and resonances. The result is that the addition of the $\it{pinched}$ kernel has only a negative impact on the results. The convergence of the PQCD contributions is not as good, and the stability region is considerably reduced. Numerically, though, the change in the final value of $m_s$(2 GeV) is well within the error given in Eq.(30), but with a narrower stability region. We must then conclude that there is no advantage in introducing an additional $\it{pinched}$ kernel in this channel.\\

Note added in proof: After this work was completed, a new and more accurate determination of $\alpha_s(M_\tau)$ by the ALEPH collaboration has been released 
\cite{ALEPH2}, which implies a narrower range for $\Lambda$, i.e. $\Lambda = 365 - 397$ MeV. Using this in our determination we obtain $m_s(2 GeV) = 99 \pm 
5 (105 \pm 6)$ MeV for $\Lambda = 397 (365)$ MeV, respectively. Combining both values gives $m_s(2 GeV) = 102 \pm 8$ MeV.

%\newpage
$\bf{Acknowledgements}$\\
We wish to thank K. Chetyrkin and A. Pivovarov for correspondence and comments.


\begin{thebibliography}{99}

\bibitem{OLD1} For a recent review see e.g. P. Colangelo, A. Khodjamirian, in: "At the Frontier of Particle Physics/ Handbook of QCD"', M. Shifman, ed. (World Scientific, Singapore 2001), Vol. 3, 1495-1576. Some recent papers include: S. Narison, Phys. Rev. D 74 (2006) 034013;
E. Gamiz, M. Jamin, A. Pich, J. Prades, F. Schwab, Phys. Rev. Lett. 94 (2005) 011803; M. Jamin, J.A. Oller, A. Pich, Phys. Rev. D 74 (2006) 074009, Eur. Phys. J. C 24 (2002) 237; K. Maltman, J. Kambor, Phys. Rev. D 65 (2002) 074013.

\bibitem{OLD2} M. Jamin, M. M\"{u}nz, Z. Phys. C 66 (1995) 633;  K.G. Chetyrkin, C.A. Dominguez, D. Pirjol, K. Schilcher, Phys. Rev. D 51 (1995) 5090; ; K. Chetyrkin, D. Pirjol, K. Schilcher, Phys. Lett. B 404 (1997) 337.

\bibitem{LATTICE} HPQCD Collaboration, C. Aubin at al., Phys. Rev. D 70 (2004) 031504; CP-PACS and JLQCD Collaborations, T. Ishikawa et al., arXiv: hep-lat/0509142, PoS LAT 2005 (2006) 057; QCDSF Collaboration, M. Gockeler at al., arXiv: hep-lat/0509159, PoS LAT 2005 (2006) 078; D. Becirevic et al., Nucl. Phys. B 734 (2006) 138; Q. Mason, H.D. Trottier, R. Horgan, C.T.H. Davies, G.P. Lepage, Phys. Rev. D 73 (2006) 114501.

\bibitem{PQCD5} P.A. Baikov, K.G. Chetyrkin, J.H. Kuhn, Phys. Rev. Lett. 96 (2006) 012003; K.G. Chetyrkin, A. Khodjamirian, Eur. Phys. J. C 46 (2006) 721;

\bibitem{PDG} Review of Particle Physics, Particle Data Group, J. Phys. G: Nucl. Part. Phys. 33 (2006) 1.

\bibitem{COND} C.A. Dominguez, N. Nasrallah, K. Schilcher, arXiv: 0711.3962 (2007).
 
\bibitem{PQCD4}
K.G Chetyrkin, A.L. Kataev, F.V. Tkachov, 
Phys. Lett. B 85 (1979) 277; M. Dine,J. Sapirstein, Phys.
Rev. Lett. 43 (1979) 668 ; W. Celmaster, R. Gonsalves {\it ibid.} 44
(1980) 560; S.G. Gorishny, A.L. Kataev, S.A. Larin, Phys. Lett. B 259
(1991) 144 ; L.R. Surguladze,M. Samuel, Phys. Rev. Lett. 66
(1991) 560; T. van Ritbergen, J.A.M. Vermarseren, S.A.
Larin, Phys. Lett. B 400 (1997) 379.

\bibitem{DAVIER} M. Davier, A. H\"{o}cker, Z. Zhang, Rev. Mod. Phys. 78 (2006) 1043.

\bibitem{CADTAU} C.A. Dominguez, K. Schilcher, J. High Energy Phys. 0701 (2007) 093.

\bibitem{CAD1} C.A. Dominguez, L. Pirovano, K. Schilcher, Phys. Lett. B 425 (1998) 193.

\bibitem{RATIO} H. Leutwyler, Phys. Lett. B 378 (1996) 313.

\bibitem {CIPT}A.A. Pivovarov, Z. Phys. C 53 (1992) 461; F. Le Diberder and A. Pich, Phys. Lett.  B 286 (1992) 147; M. Jamin, J. High Energy Phys. 0509 (2005) 058.

\bibitem{MALT} K. Maltman, Phys. Lett. B 440 (1998) 367; C.A. Dominguez, K. Schilcher, Physics Letters  B 448  (1999) 93; 581 (2004) 193.

\bibitem{ALEPH2} M. Davier, S. Descotes-Genon, A. H\"{o}cker, B. Malaescu, Z. Zhang,  arXiv:0803.0979. 


\end{thebibliography}
\end{document}